\documentclass[12pt]{article}

\usepackage{times,color}
\usepackage{amsmath}
\usepackage{subfigure,epsfig,graphicx,epstopdf}
\usepackage[colorlinks]{hyperref}
\hypersetup{linkcolor = {blue},	citecolor = {blue},	urlcolor = {black}}
\usepackage{cite}
\usepackage{upgreek}

\newenvironment{sciabstract}{%
\begin{quote} \bf}
{\end{quote}}

\newcounter{lastnote}

\begin{document}

\title{Femtojoule-per-operation photonic computer for the subset sum problem}
\author{Tian-Yu Zhang$^{1,2,3}$$^{\dagger}$, Xiao-Yun Xu$^{1,2}$$^{\dagger\ast}$, Wen-Hao Zhou$^{1,2}$, Xiao-Wei Wang$^{1,2}$,\\ Chu-Han Wang$^{1,2}$, Yi-Jun Chang$^{1,2}$, Ying-Yue Yang$^{1,2,3}$, Jie Ma$^{1,2}$, \\Ka-Di Zhu$^{3 \ast}$, Xian-Min Jin$^{1,2,4\ast}$\\
\normalsize{$^1$Center for Integrated Quantum Information Technologies (IQIT),}\\
\normalsize{School of Physics and Astronomy and State Key Laboratory of Advanced} \\
\normalsize{Optical Communication Systems and Networks,}\\
\normalsize{Shanghai Jiao Tong University, Shanghai 200240, China}\\
\normalsize{$^2$CAS Center for Excellence and Synergetic Innovation Center in Quantum}\\ \normalsize{Information and Quantum Physics,}\\
\normalsize{University of Science and Technology of China, Hefei, Anhui 230026, China}\\
\normalsize{$^3$Key Laboratory of Artificial Structures and Quantum Control (Ministry of Education)}\\
\normalsize{and School of Physics and Astronomy,}\\
\normalsize{Shanghai Jiao Tong University, Shanghai 200240, China}\\
\normalsize{$^4$TuringQ Co., Ltd., Shanghai 200240, China}\\
\normalsize{$\dagger$ These authors contributed equally to this work.}\\
\normalsize{$^\ast$E-mail: xiaoyunxu@sjtu.edu.cn}\\
\normalsize{$^\ast$E-mail: zhukadi@sjtu.edu.cn}\\
\normalsize{$^\ast$E-mail: xianmin.jin@sjtu.edu.cn}
}

\date{}

\baselineskip 24pt

\maketitle

\begin{sciabstract}
Energy-efficient computing is becoming increasingly important in the information era. However, electronic computers with von Neumann architecture can hardly meet the challenge due to the inevitable energy-intensive data movement, especially when tackling computationally hard problems or complicated tasks. Here, we experimentally demonstrate an energy-efficient photonic computer that solves intractable subset sum problem (SSP) by making use of the extremely low energy level of photons ($\sim\textbf{10}^{\textbf{-19}}$ J) and a time-of-flight storage technique. We show that the energy consumption of the photonic computer maintains no larger than $\textbf{10}^{\textbf{-15}}$ J per operation at a reasonably large problem size N=33, and it consumes $\textbf{10}^{\textbf{8}}$ times less energy than the most energy-efficient supercomputer for a medium-scale problem. In addition, when the photonic computer is applied to deal with real-life problems that involves iterative computation of the SSP, the photonic advantage in energy consumption is further enhanced and massive energy can be saved. Our results indicate the superior competitiveness of the photonic computer in the energy costs of complex computation, opening a possible path to green computing.\\

\end{sciabstract}


\noindent As excessive consumption of energy heavily burdens the society\cite{portegies2020ecological,khanna2021multi}, it has become a global consensus to alleviate energy consumption to achieve carbon emission peak and carbon neutrality. A large proportion of worldwide energy is devoted to daily processing of mega data\cite{masanet2020recalibrating}. As the most widely used tool to process information, conventional electronic computer tends to become ``energy-hungry beast" in the context of complex iterative computation. Even the most energy-efficient supercomputer could easily consume several megawatts of electric power, which is sufficient to sustain a small city\cite{feng2007green500}. Moreover, the accompanying excessive heat could reduce system reliability since computing at higher temperature is error-prone, which also hinders the computational power development of electronic computer\cite{valentini2013overview}. Though a myriad of optimization strategies, ranging from hardware to software \cite{valentini2013overview, barroso2007case,rong2016optimizing}, have been proposed, the data movement bottleneck steaming from von Neumann architecture radically hampers the further enhancement of electronic computers in energy efficiency\cite{lee2021energy}. Hence it is in urgent need to develop a novel energy-efficient computing regime.

Nondeterministic-polynomial-time (NP)-complete problems are a class of tasks intractable to solve on electronic computers, since the solution space grows superpolynomially with the increase  of problem size\cite{garey2002computers, xu2020scalable}. As a consequence, enormous computing resources are required when solving NP-complete problems. Also, various realistic scenarios such as bioinformation\cite{lynce2006efficient,eghdami2021ssa}, electronic design\cite{marques2000boolean}, finance\cite{orus2019quantum,biesner2022solving}, medical diagnosis\cite{arle2021medical} and cryptosystem\cite{murakami2018security,murakami2012public} are related to NP-complete problems. The computational hardness and broad application reveal that NP-complete problems could be metrics for evaluating novel computing paradigms\cite{xu2020scalable,zhang2021quantum,vazquez2018optical,braich2002solution,centrone2021experimental}. Diverse non-von Neumann computing paradigms such as neuromorphic computing\cite{demasius2021energy,chen2016eyeriss,pei2019towards}, quantum computing\cite{ikonen2017energy,elsayed2019review,britt2017quantum}, biocomputing\cite{nicolau2016parallel,adleman1994molecular} and optical computing\cite{nakajima2021scalable,xu2023integrated,wang2022optical,lee2021energy} have made some progress in energy efficiency. However, these achievements are generally limited to
specific tasks, computation precision, and there is still a lot of room for improvement. It is worth mentioning that the inherent ultralow energy of single photons ($\sim 10^{-19}$ J) provides an opportunity to build a photonic computer with remarkable energy efficiency\cite{wang2022optical}. Moreover, the state-of-the-art single-photon sources\cite{uppu2020scalable,zhong2020quantum} and commercially available single-photon detectors \cite{yin2021improving,you2020superconducting} allow to process information at single-photon level, thus providing a possibility to achieve complex computation with only a few photons consumed.

Here we present an energy-efficient photonic computer for the subset sum problem (SSP), a typical NP-complete problem. We map the SSP to a three-dimensional waveguide network, which is embedded in the photonic chip fabricated by femtosecond laser direct writing technique. Heralded single photons are injected into the SSP waveguide network to carry out the computation. An array of avalanche photodetectors (APDs) is used to detect the photons, whose arriving time are recorded by the time-of-flight storage (ToFS) module. Based on the time information, the so-called timestamp, we reconstruct the probability distribution of photons at the output ports of the waveguide network, revealing the solution of the SSP. We also analyze the reliability and energy-consumption performance of the photonic computer with the growth of problem size and the increase of the execution count of the SSP, respectively, aiming to show its energy advantage from different aspects.

\section*{Results}
\subsection*{Architecture of the photonic computer}
\begin{figure*}
\centering
\includegraphics[width=0.95\columnwidth]{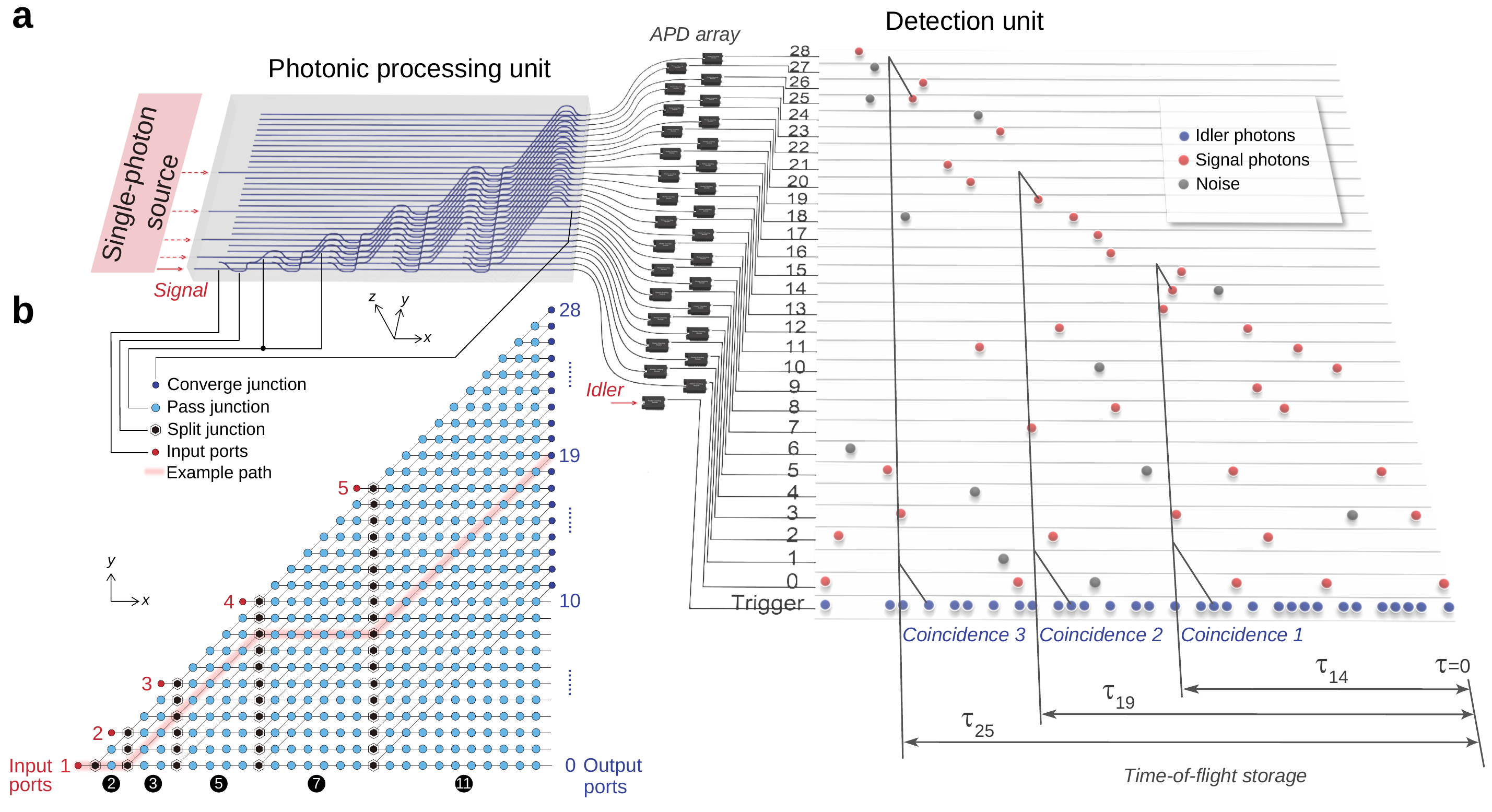}
\caption{\textbf{Architecture of the photonic computer.} \textbf{a}, The photonic processing unit (PPU) is embedded with a waveguide network, where the computation of various SSPs are activated by injecting signal photons into the PPU through different input ports, as the red arrows denoted on the left. After being emitted from the PPU, photons are detected by an array of APDs, and then the successful detection is transmitted to a time-of-flight storage module to record the arriving time (the so-called timestamp) of photons. The appearance of signal photons should be validated by the twofold coincidences between them and idler photons acting as triggers, whose timestamp $\tau_{i}$ will be used to retrieve the computing results of the SSP. \textbf{b}, The abstract version of the waveguide network in \textbf{a}, which explains the computation process of the SSP on the photonic computer. Split junctions transmit photons to two directions with the same probability, and pass junctions guide photons to propagate along their original direction. Converge junctions are used to gather photons from different directions. The inclusion of an element in the summation or not depends on whether photons have a displacement in the vertical direction (i.e., photons propagate diagonally), and the value of elements are denoted by the integers below the network. The serial number of the output ports give all the possible subset sums. An example path highlighted in pink shows that elements 3, 5, and 11 are included, giving rise to a subset sum 19.}
\label{fig1}
\end{figure*}

The photonic photonic computer for the SSP contains three parts, the heralded single-photon source (see Supplementary Section \uppercase\expandafter{\romannumeral1}), the photonic processing unit (PPU) and the detection unit composed of an array of APDs and a ToFS module. As displayed in Fig. \hyperref[fig1]{1a}, signal photons from the heralded single-photon source are injected into the PPU to perform the computation, and idler photons act as triggers. The photons are detected by the APDs and their arriving time is recorded by the ToFS module whose 30 channels (1 trigger channel and 29 detection channels) are connected to the APDs one to one. We validate the appearance of signal photons by the coincidences between them and idler photons. This post-selection process (see Methods) helps to distinguish the signal photons from noise which stems from the ambient environment and dark counts of the detectors. Three coincidence examples for channel 14, 19, and 25 are marked. The arriving time of the idler photons in the coincidence events is considered as the timestamps, which are denoted as $\tau_{i}$ where ${i}$ is channel number. The computation results of the SSP can be retrieved relying on the timestamps, as shown in the next section.

As the heart of the photonic computer, the PPU plays a similar role to the CPU in electronic computers. The architecture of the PPU is modified from the work of Xu \textit{et al.} \cite{xu2020scalable}, with a more compact and efficient structure, as well as reduced insertion loss. As shown in Fig. \hyperref[fig1]{1a}, a three-dimensional waveguide network (the dark blue lines) is embedded in the PPU. Photons propagate in the waveguide network to search for the solution of the SSP. The network is built with standardized functional modules, i.e., split junctions, pass junctions, and converge junctions, as demonstrated in Fig. \hyperref[fig1]{1b}. Split junctions (denoted by hexagonal nodes) take charge of transmitting photons to the horizontal (\textit{x}-direction) and diagonal paths with equal probability. The sky blue circular nodes are pass junctions that are used to guide photons to propagate in their original direction. The dark blue circular nodes are converge junctions which are responsible for gathering photons from different directions and transmitting them to the \textit{x}-direction paths. The specific physical structure of functional modules are depicted in Supplementary Section \uppercase\expandafter{\romannumeral2}, and the fabrication details can be found in Methods. 

Given a set $S$ of integers and a target $T$, the SSP asks whether there exists any subset of $S$ whose sum is equal to $T$ \cite{karp1972}. Once the signal photons are launched into the network through the first input port (see chip packaging in Methods), the computation process for solving the SSP instance where $S=\{2, 3, 5, 7, 11\}$ is activated. Note that there are several input ports, as denoted by the red arrows in Fig. \hyperref[fig1]{1a} and the red circular nodes in Fig. \hyperref[fig1]{1b}. Different choices of input ports correspond to the computation of different SSP instances. As presented in Fig. \hyperref[fig1]{1b}, the incident photons first encounter a split junction. Afterwards, they go through several pass junctions diagonally or horizontally, which represent including an element of $S$ into the summation or not, respectively. Each row of split junctions can be seem as a block, and the number of pass junctions between adjacent split junction blocks indicates the value of the elements, as denoted by the integers below the network. We classify such sandwiched regions as blocks of pass junctions. Apparently, the number of the two kinds of blocks are equivalent, and is equal to the number of elements in the set $S$. After going through all the blocks, photons arrive at the converge junctions and exit from the output ports whose serial number represent the corresponding sums. For instance, the path highlighted in pink demonstrates the summation including elements 3, 5, and 11, and the sum is 19. In contrast to sequentially-operating electronic computers which enumerate all the subsets temporally, we map the solution space of the SSP to a physical embodiment which encodes all the results spatially. As aforementioned, in our photonic computer, the spatial position of the output ports represents all the possible subset sums. 

\subsection*{Retrieved computing results based on the timestamp of photons}

\begin{figure*}
\centering
\includegraphics[width=0.95 \columnwidth]{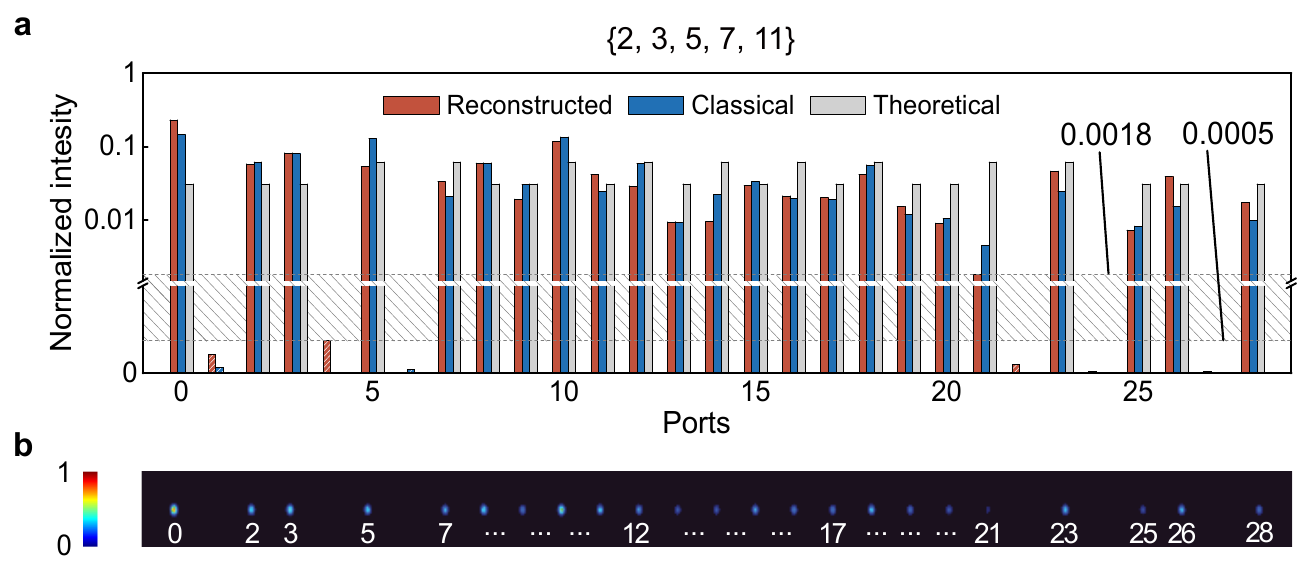}
\caption{\textbf{The distribution of photons at the output ports in the SSP case \{\textbf{2, 3, 5, 7, 11}\}}. \textbf{a}, The probability distribution of photons is reconstructed by the timestamp of the fifth photon arriving at each detection channel. The theoretical intensity distribution is calculated based on an ideal model, and the evolution of 780 nm coherent light in the waveguide network results in the classical intensity distribution. The threshold applicable in our experiments has a proper tolerance interval, which is revealed by the upper bound (0.0018) and lower band (0.0005) of the band filled with solidus pattern. \textbf{b}, The visualization of the recontructed probability distribution of photons of the SSP case \{2, 3, 5, 7, 11\}, as plotted by a line of Gaussian spots.}
\label{fig2}
\end{figure*}

According to the definition of the SSP, a nonzero probability of photons appearing at certain output port indicates a solution of the SSP. Namely, in this case the answer to the problem is YES and the subset sum $T$ is denoted by the serial number of the output port. Otherwise, the answer is NO. In general, an earlier detection of photons at the output port suggests a larger probability. Therefore, the timestamp of the injected photons can be exploited to reconstruct the probability distribution at the outputs, as described in Eq.\ref{eq:p}\cite{zhou2022timestamp}, where $\tau_{i,n}$ is the timestamp of the $n^{th}$ photon detected at output port $i$, and $p_{i,n}$ is the normalized probability at output port $i$. 
\begin{equation} \label{eq:p}
p_{i,n}=
\frac{\tau_{i,n}^{-1}}{\sum_{i}\tau_{i,n}^{-1}}
\end{equation}
Based on the equation, we reconstruct the probability distribution of the specific instance where $S = \{2, 3, 5, 7, 11\}$. Fig. \hyperref[fig2]{2a} exhibits the results when we utilize the timestamp of the fifth detected photons. 

Ideally, the probability at particular ports which correspond to none subset sum should be zero. As displayed in Fig. \hyperref[fig2]{2a}, the theoretical model gives rise to only three values, i.e., 0.0625, 0.03125, and 0. Zero intensity indicates that a sum does not exist. Otherwise, it exists. However, in light of unavoidable noise, there could be coincidence events at those ports. To identify channels with validated signal (i.e., photon appearance) from noise-only channels, we can legitimately set a threshold. The probabilities beyond the threshold are considered as the existence of subset sums, while those below the threshold (highlighted with white solidus pattern) represent the absence of subset sums. The threshold is set as $1/10$ of the detected probability at output port 28, where the weakest validated signal is supposed to appear since it corresponds to the subset containing all the elements and the path to output port 28 is the longest. Following the above rules, the retrieved computing results of the SSP are in accordance with those of the theoretical model, verifying the accuracy of the photonic computer is 100\%, which means the solution is deterministic. Note that more options for the threshold are available, and the tolerance band of the threshold is indicated by the band filled with solidus in Fig. \hyperref[fig2]{2a}. The upper bound (0.0018) and the lower bound (0.0005) of the threshold depend on the minimum signal and the maximal noise, respectively. The good signal-to-noise ratio (SNR, equal to 0.0018 divided by 0.0005) results in a reasonably large tolerance band. It should be noticed that a higher SNR can be attained with an employment of the timestamp of later photons, i.e., $n\textgreater5$ (Supplementary Section III). For visualization, we plot the reconstructed probability distribution as a line of Gaussian spots, as shown in Fig. \hyperref[fig2]{2b}. The appearance of a spot demonstrates that there exists at least one subset whose sum is equal to the integer, while the absence indicates the opposite. Besides, the computation of various SSP cases can be achieved by switching the input port, and all the retrieved computing results are highly accurate (Supplementary Section IV).

In addition to intrinsic high performances of the PPU, the reliability of the timestamp-based protocol is also a crucial guarantee of the high accuracy. Therefore, we further investigate the influence of the timestamp-based protocol through a comparison with purely classical case where a beam of 780 nm coherent laser into the waveguide network and the intensity distribution is directly captured by a charge-coupled device (CCD). As presented in Fig. \hyperref[fig2]{2a}, the classical intensity distribution possesses similar characteristics to the reconstructed case. More specifically, the similarity between the results in the two protocols $\eta=\sum_{i}\sqrt{p_{i,5}p_{i}^{\prime}}$ \cite{zhou2022timestamp} is up to 97.4 $\%$ (the normalized intensity distribution is $p_{i}^{\prime}$ in the classical case), which also benefits from the high resolution 64 ps of the ToFs. In other words, the probability distribution of photons, as inferred from the timestamp of the fifth photon, accurately reflects the realistic evolution of millions of photons. Therefore, it elucidates the ease with which computing results of 100\% accuracy are attainable by establishing a suitable threshold that supports a wide tolerance range.

\subsection*{Energy expenditure}
Energy efficiency is a critical criterion to evaluate the performance and sustainability of computers. Here we investigate the energy-consumption performance of the photonic computer when it solves the SSP instance where $S$ consists of the first $N$ primes. Representative CPU, graphics processing unit (GPU), and the most energy-efficient supercomputer are compared with the photonic computer. Note that the energy consumption of the photonic computer is related to the number of photons required to correctly reconstruct the probability distribution. In the proposed timestamp-based protocol, we are able to achieve the goal with only a few photons detected at the output ports, rather than millions of photons (estimated by the minimum power density of the CCD, LBP2-HR-VIS2, Newport), as demonstrated in the above section. 

\begin{figure*}
\centering
\includegraphics[width=0.95 \columnwidth]{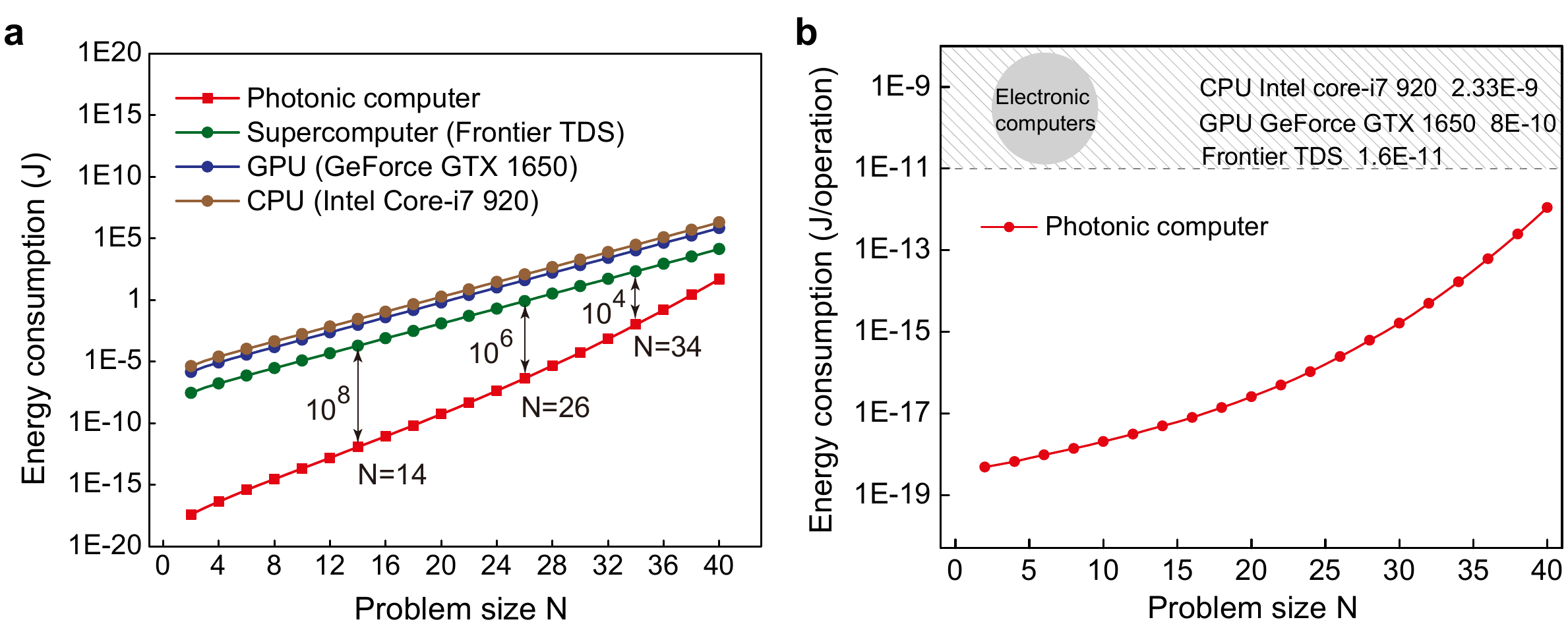}
\caption{\textbf{Energy-consumption performance.} \textbf{a}, The total energy consumption of solving the SSP with successive primes \{2, 3, 5, 7, 11, ...\} on representative CPU, GPU, supercomputer and the photonic computer under the timestamp-based protocol using the fifth photons. The photonic computer outperforms its electronic competitors with remarkable advantages, which are maintained as the problem expands. \textbf{b}, The energy costs of a single operation for fully-fledged electronic computers and the photonic computer. The photonic computer consumes the energy of about a single photon ($\sim10^{-19}$ J) at a relatively small problem size, and remains no larger than the level of $10^{-15}$ J even when the problem size N increases to 33.}
\label{fig3}
\end{figure*}

We define the total photon loss as the energy consumed by the photonic computer, which is determined by the propagation loss of the waveguide network, the detection efficiency and the heralding efficiency of the single-photon source (see Methods). Details on the energy costs of the electronic computers can also be found in Methods. As illustrated in Fig. \hyperref[fig3]{3a}, the photonic computer consumes up to $10^{8}$ times less energy than the electronic rivals at a medium size of $N=14$, even compared with the supercomputer. The photonic superiority remains considerable as the problem size $N$ grows, with $10^{6}$ and $10^{4}$ times less energy-consumption than the electronic competitors at $N=26$ and $N=34$, respectively. From the aspect of single operation, the photonic computer surpasses fully-fledged electronic computers by a significant order of magnitude. As presented in Fig. \hyperref[fig3]{3b},the energy consumed by the photonic computer can be down to single-photon level ($\sim10^{-19}$ J) when $N\textless7$, and no larger than the level of $10^{-15}$ J when $N\leq33$. It strongly shows the superiority of our photonic computer in a single calculation of SSP.

\begin{figure*}
\centering
\includegraphics[width=0.7 \columnwidth]{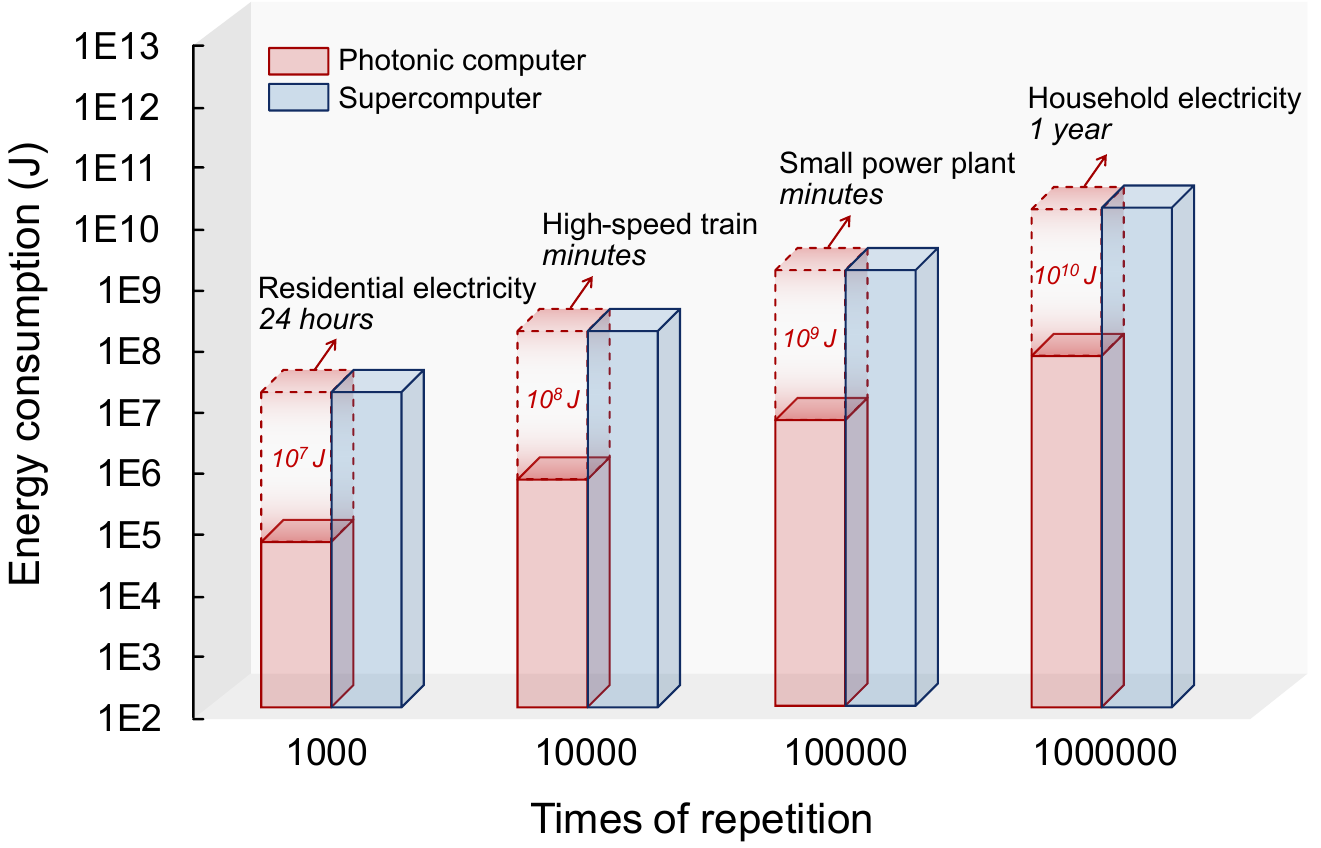}
\caption{\textbf{Energy-consumption for realistic scenarios involving repetitive computation of the SSP.} In the case of realistic applications where the computation of the SSP is the critical step, computers have to compute the SSPs in a iterative manner. For instances where $S$ includes 40 successive primes, the energy-consumption gap between the photonic computer and the supercomputer continues to widen, as the times of repetition increases. Compared to the supercomputer, the photonic computer can save energy enough to support large amounts of social power use, as illustrated above each column.}
\label{fig4}
\end{figure*}

Although the current photonic computer cannot outperform supercomputers in every aspect, particularly for very large problem sizes (where $N \geq 47$), it offers significant energy savings in scenarios where they remain more energy-efficient. More specifically, the energy advantage of our photonic computer is further strengthened in realistic applications where iterative computation of SSP are involved. As shown in Fig. \hyperref[fig4]{4}, we estimate the energy costs on the photonic computer and supercomputer for workload balancing problem (a typical real-life problem with NP-hard complexity)\cite{schwerdfeger2016fast}. Given several identical parallel machines and independent jobs with integer processing times, the workload balancing problem asks for the most balanced assignment to minimize the normalized sum of squared workload deviations. Compared with traditional algorithm, the problem can be solved more efficiently by an SSP-based heuristic algorithm, where the assignment is iterated to reach an optimal solution and the computation of  the SSP instances during each iteration represents the validation of the assignment (Supplementary Section \uppercase\expandafter{\romannumeral5}). It is foreseeable that the energy-consumption advantage of the photonic computer builds up during such repeated computations. As demonstrated in Fig. \hyperref[fig4]{4}, when it comes to N = 40, the gap of the energy consumption between the photonic computer and the supercomputer is increasingly widening, as the times of repetition increases. It is worthing stressing that the gaps amount to huge expenses in social power use, such as one-year household electricity, as denoted in the figure.

\section*{Discussion}
In conclusion, we present an energy-efficient photonic computer to solve the SSP, a NP-complete problem, and experimentally investigate its reliability and energy-consumption performance. Sophisticated femtosecond laser writing technique is applied to fabricate the PPU with embedded three-dimensional waveguide networks, in which photons propagate to execute computation tasks\cite{xu2021quantum,gao2022quantum}. In our protocol, we reconstruct the probability distribution at the output ports of the PPU based on the arriving time of photons, and retrieve the computing results of the SSP according to the probability distribution. Even against the best supercomputer, our photonic computer maintains several orders of magnitude of energy advantages, and energy consumption for a single operation can be as low as femtojoules. The photon-enabled energy advantage not only confirms the superiority of non-von Neumann computers, but also shows the competitive of photonic computing.

The key of our approach to reach ultrahigh energy-efficiency is that the probability distribution of photons is inferred by the timestamp of the first few photons, free of accumulating a mass of photons to form the intensity distribution. We show that, with the timestamps of the fifth photons, the retrieved computation results of various SSP instances are consistent with those in theoretical and classical scenarios. Namely, only five photons need to detect at each output port in our timestamp-based protocol, which is much less than that of the classical scenario where a bunch of photons is required. 
The excellent performance is also assisted by state-of-the-art single-photon detectors with ultralow dark count rate. In addition, as a non-von Neumann computer, our photonic computer gets rid of the huge energy consumption caused by data movement, and avoids the complicated and extremely energy-intensive cryogenic refrigeration system. The last but not least, the natural low detectable energy level of photons underpins the photonic computer to be an outstanding green computer.

In principle, given the polynomial-time reducibility among NP-complete problems\cite{ruiz2011survey}, a multitude of these problems can be efficiently solved on this photonic computer, which is also benefit from the programmability of such photonic computer\cite{xu2024reconfi}. This could greatly relieve the energy pressure in solving such a large class of intractable problems, and with deterministic solution. In addition, exploring such special-purpose hardware could ease the burden for universal optical computing, where in general resolving the enormous energy dissipation of optical-to-electronic conversions and interconnects is complicated and hard\cite{miller2010optical,miller2017attojoule}. More broadly, the unique and reliable timestamp-based approach used to derive probability distribution results also open the way to further applications in other tricky tasks based on photonic hardware, such as optical neural networks\cite{shen2017deep}, quantum transport\cite{xu2021quantum}, and topological photonics\cite{noh2020braiding}, thereby inspires novel energy-saving measures.

\section*{Methods}
\noindent \textbf{Post selection of photons.} 
Extraction of twofold coincidences should consider an appropriate coincidence time window and the relative time delay between signal and idler photons. The delays are variable for different detection channels, since the signal photons go through paths with different length. Specifically for each detection channel, we scanned the delay from -1 ns to 3 ns to obtain the corresponding coincidence counts. Normally, the curve of coincidence count versus delay is of Gaussian shape, and the peak corresponds to the actual delay of the detection channel. In terms of detection channels where counts are rare and Gaussian peaks are hard to find, all the coincidence counts could be noise. We analyzed about 5 million timestamps collected in our experiments, and found that Gaussian peaks clearly appear at counts-rich (from 79 to 12068) channels while does not appear at non-rich (from 0 to 11) channels. For the latter case, it makes no sense to determine the actual delays according to the peak coincidence counts, and thus we set the delay for these channels to 1 ns (the most common value among the delays in counts-rich channels). After calibrating the time delays, we filtered out all the photon pairs within 1 ns coincidence window.\\

\noindent \textbf{Fabrication of the 3D waveguide network.} High-performance integrated devices are the foundation to guarantee the on-chip computation and achieve desired computing results, and also critical to the scalability of the PPU. Ideally, the computation of the SSP requires that the splitting ratio of split junctions is exactly 50:50, and the light in the two waveguides of pass junctions is totally decoupled and can be successfully merged together in converge junctions without any energy loss. Experimentally, high-quality implementation and modulation of the above parameters rely on the femtosecond laser writing technique. We fabricated waveguides via a 513 nm femtosecond laser, of which the pulse duration is 290 fs and the repetition rate is 1 MHz. The laser was locked by a beam-pointing stabilizer to improve the fabrication precision. A borosilicate substrate (Corning Eagle XG glass) was placed on the three-dimensional translation stage with moving speed of 10 mm/s. Eventually the laser was focused by a 50$\times$ objective (NA=0.55) into the glass at a depth of 132.5 $\upmu$m to inscribe the waveguide network, with a cylindrical lens to compensate for its aberration in the glass.\\  

\noindent \textbf{Chip packaging.} The incident photons are coupled into the waveguide network through a single-mode V-groove fiber array, which has 30 channels with a spacing of 50 $\upmu$m (30ch 0D 50P FA-FC/UPC, Zhongshan Meisu Technology Co.,Ltd.). The fiber array is also used to delivery the emitted photons from the output ports to the APDs. The perfect alignment between the fiber array and waveguide network ports is the key to achieve the highest coupling efficiency. Position calibration was performed on a 6-axis nano-positioning flexure stage with high precision adjustment. When the optical power of the first and last port reach their maximum, the best alignment is realized. After that we moved the fiber array in \textit{x}-direction so that it is as close to the chip as possible, and gummed them together by an optical glue (Norland Optical Adhesive 61, Norland Products).\\

\noindent \textbf{Energy-consumption estimation.} To figure out the total energy consumption of the photonic computer, we first calculated the minimal number of photons required to ensure the success of the computation, and then acquire the total energy consumption through multiplying it by the propagation loss (in the form of percent) of the waveguide network. In our timestamp-based protocol using the fifth photons, at least five photons need to detect at the output port of the longest path. The minimal number of incident photons is equal to five divided by the product of heralding efficiency, the transmissivity of the longest path and the detection efficiency of APD. Note that by the time the APD connected to the longest path detects the fifth photon, the others have collected more than five photons due to less photon loss in the shorter paths. In more detail, the transmissivity of the longest path is calculated on the basis of photon attenuation in waveguides, actual splitting ratio of split junctions, and the loss of converge junctions. Considering that we have prepared several networks and selected the optimal one according to their intensity distributions, it is reasonable to assume the imbalance of the virtual splitting ratio is within 5 $\%$. In addition, the measured transmissivity of the converge junctions is 52 $\%$. Finally, we can further obtain the energy consumption of a single operation, which was calculated by dividing the energy of lost photons in the longest path in our waveguide network by the number of elements in $S$.

Conventional electronic computers adopt sequential operation. Given a set $S$ with $N$ elements, to verify whether there exists a subset sum equivalent to target $T$, the computer first reads all the elements from the memory to the CPU. For the first element, there are two options, i.e., to be included in the subset or not, thus forming two probable subsets. After that, the electronic computer performs the same operation on the second element in the set $S$ and forms four subsets. For all the $N$ elements we eventually get $2^{N}$ subsets. Under this condition, the total energy consumption of electronic computers includes three parts, the energy consumed to read all elements from the memory to the CPU, the energy of data movement in the process of generating new possible subsets, and the energy of floating point operations, which is used to calculate the sum of each subset. The third part is assessed through dividing the number of required addition operations by floating point operations per second per watts (i.e., performance per watt, a metrics of energy efficiency of electronic computer hardware). On the Green500 list, the most energy-efficient supercomputer consumes $1.6\times10^{-11}$ J per floating point operation\cite{Green500List}. The same parameter for the CPU and the GPU refer to their thermal design power (TDP) and performance\cite{CPU,GPU}. According to the quantitatively evaluated energy costs for data movement between the memory and the CPU, which is approximate 100 times more than those for single floating point operations\cite{kestor2013quantifying}, we estimated the first and the second part of energy consumption. Based on the energy costs in computation, we also estimated the energy used for the cooling system, which is about three quarters of the former\cite{feng2007green500}, and added it to the total energy consumption of electronic computers.

It should be noted that in our energy consumption estimates, the energy required for external equipment and the fabrication process are not included. This is because comparing these aspects between photonic and electronic computers is complex and challenging to equate fairly. The construction of electronic computers involves energy-intensive steps such as lithography, etching, and deposition, which do not always yield fully functional chips. This is in addition to the energy used in assembly, testing, maintenance, and the extensive cooling systems required. In contrast, the energy consumption for fabricating our photonic computer is likely much lower than that of a supercomputer or even typical electronic computers, largely dependent on the energy usage of the lasers used in chip fabrication. However, it is still too complicated to estimate.

\subsection*{Data availability}
The data that support the findings of this study are available from the corresponding authors on reasonable request.

\subsection*{Acknowledgements}
The authors thank Jian-Wei Pan for helpful discussions. This research is supported by the National Key R\&D Program of China (2019YFA0308700, 2019YFA0706302 and 2017YFA0303700); National Natural Science Foundation of China (NSFC) (12104299, 61734005, 11761141014,
11690033); Science and Technology Commission of Shanghai Municipality (STCSM)(20JC1416300, 2019SHZDZX01); Shanghai Municipal Education Commission (SMEC)(2017-01-07-00-02-E00049); China Postdoctoral Science Foundation (2021M692094); China Postdoctoral Science Foundation (2022T150415); Natural Science Foundation of Shanghai
(No. 20ZR1429900). X.-M.J. acknowledges additional support from a Shanghai talent program and support from Zhiyuan Innovative Research Center of Shanghai Jiao Tong University.

\subsection*{Author contributions} 
X.-M.J. and K.-D.Z. conceived and supervised the project. T.-Y.Z., X.-Y.X., W.-H.Z., X.-W.W. C.-H.W., Y.-J.C., Y.-Y.Y. and J.M performed the experiment and analyzed the data. X.-Y.X. and T.-Y.Z. prepared the photonic processing unit. T.-Y.Z., X.-Y.X. and X.-M.J. wrote the paper, with input from all the other authors.

\subsection*{Competing interests}
The authors declare no competing interests.

\baselineskip21pt

%

\end{document}